\documentclass{article}
\usepackage[psamsfonts]{amssymb}
\usepackage{amsmath}
\usepackage{cite}
\def\Ref#1{(\ref{#1})}
\usepackage{cite}
\def\d{\mathrm{d}}
\def\ri{\mathrm{i}}
\def\c{\mathrm{c}}
\def\rc{\mathrm{cl}}
\def\rw{\mathrm{w}}
\def\rs{\mathrm{s}}
\def\m{\mathrm{m}}
\author{L. Lavaei-Yanesi\footnote{yalda57l@yahoo.com}\ \ and
M. Khorrami\footnote{mamwad@mailaps.org}
\\ {\small Department of Physics, Alzahra University, Tehran
1993891167, Iran.}}
\title{Phase structure of the quartic-cubic generalized two
dimensional Yang Mills U($N$) on the sphere}
\date{}
\begin{document}
\maketitle


\begin{abstract}
\noindent The large-$N$ behavior of the quartic-cubic generalized
two dimensional Yang Mills U($N$) on the sphere is investigated,
for small cubic couplings. It is shown that single transition at
the critical area which is present for the quartic model, is split
into two transitions, both of them are third order. the phase
diagram of the system for small cubic couplings is obtained.
\end{abstract}
\section{Introduction}
During recent years, there has been extensive studies of the
two-dimensional Yang-Mills theory (YM$_2$) and generalized
Yang-Mills theories (gYM$_2$'s)
\cite{1,2,3,4,5,6,7,b,c,10,9,19,12,12p,13a,27,d,20}. A review on
this topic is \cite{aba}. These are important integrable models
which are expected to shed light on some basic features of pure
QCD in four dimensions. Besides, there are certain relations
between these theories and string theories. To be more specific,
there relations are between the large-gauge-group limit of YM$_2$
and string theory. For example, it is shown in \cite{3}, \cite{6},
and \cite{7}, that a gauge theory based on SU($N$) is split at
large $N$ into two copies of a chiral theory, encapsulating the
geometry of the string maps. The chiral theory associated to the
Yang--Mills theory on a two--manifold $\cal M$ is a summation over
maps from the two--dimensional world sheet (of arbitrary genus) to
the manifold $\cal M$. This leads to a $1/N$ expansion for the
partition function and observables, which is convergent for all of
the values of area$\times$coupling constant on the target space
$\cal M$, if the genus is one or greater.

The partition function and the expectation values of the Wilson
loops of YM$_2$ were obtained for theories on a lattice \cite{1,8}
and continuum \cite{4,9,19}. The partition function and the
expectation values of the Wilson loops of gYM$_2$'s were
calculated in \cite{c,10}. All of these results are in terms of
summations over the irreducible representations of the
corresponding gauge group. When the group is large, these
summations are dominated by some specific representations (the so
called classical representations) in some cases, and one can
obtain closed-form expressions for that representation and the
observables of the theory.

The large-$N$ limit of U$(N)$ YM$_2$ on a sphere was first studied
in \cite{11}. There the classical representation was calculated
and it was shown that the free energy of theory has a logarithmic
behavior with respect to the area of the sphere. In \cite{12}, it
was shown that that behavior is correct provided the area of the
sphere is smaller than some critical area, and it was further
shown that at that area a third order phase transition occurs.
This transition is similar to the well-known Gross-Witten-Wadia
phase transition for the lattice two dimensional multicolor gauge
theory \cite{13,14}. The phase structure of the large-$N$ YM$_2$,
generalized YM$_2$'s, and nonlocal YM$_2$'s on a sphere were
further discussed in \cite{27,20,37,39}. The large-$N$ limit of
the partition function of YM$_2$ on orientable compact surfaces
with boundaries was discussed in \cite{AK}, and the large-$N$
behavior of Wilson loops of YM$_2$ and gYM$_2$ on sphere were
investigated in \cite{l}. The critical behaviors of these
quantities have been also studied.

A gYM$_2$ with the gauge group U($N$) on a sphere is characterized
by a function $G$, as introduced in \cite{20}, for example. Most
of these investigations on gYM$_2$'s have been on theories with
even $G$. In fact, as it was pointed out in \cite{20}, if $G$ is a
polynomial then its degree should be even in order that the
partition function be convergent. So if one takes $G$ a monomial,
it should be a monomial of even degree. So to study theories based
on non-even $G$'s, the corresponding $G$'s should contain at least
two terms. A quadratic-linear combination gives nothing new, as
the linear term can be absorbed in a shifted variable. So the
simplest nontrivial non-even $G$ must contain a quartic term.

In this paper the behavior of a gYM$_2$ based on the gauge group
U($N$) on a sphere is studied, for which the function $G$ is a
binomial containing a quartic term as well as a cubic one. The
study is performed for small values of the cubic coupling. In
section 2 the behavior of this model for areas smaller than the
critical area (the weak phase) is investigated and compared with
the behavior of a theory without the cubic term. In section 3 the
transitions properties at the first transition area  are
investigated. In section 4 it is seen that there exists another
transition, and the relation between this transition to the first
one is investigated. Section 5 is devoted to the concluding
remarks.
\section{The weak phase}
To fix notation, let us first quickly review the expression for
the partition function of a U($N$)-$gYM_2$ on the sphere, in the
large $N$ limit \cite{20}.

The partition function of a $gYM_2$ on the sphere of area $A$ is
\begin{equation}\label{dp1} Z=\sum_r d_r^2 e^{-A \Lambda(r)},
\end{equation} where $r$'s label
the irreducible representations of the gauge group, $d_r$ is the
dimension of the representation $r$, and $\Lambda$ is a
combination of the Casimirs of the group. For the gauge group
U$(N)$, one can take $\Lambda$ as:
\begin{equation}\label{dp2}
\Lambda(r)=\sum_{k=1}^p \frac{a_k}{N^{k-1}}\,C_k(r),
\end{equation}
in which $C_k$ is the $k$'th Casimir of the group, and $a_k$'s are
arbitrary constants. The representations of the group U$(N)$ are
parametrized by $N$ integers $n_1\geq n_2\geq ...\geq n_N$. In
terms of this parametrization, one has
\begin{align}\label{dp3}
d_r=&\prod_{1\leq i< j\leq N} \left(1+\frac{n_i-n_j}{j-i}\right),
\nonumber\\
C_k=&\sum_{i=1}^{N} [(n_i+N-i)^k-(N-i)^k].
\end{align}
In order that the partition function \Ref{dp1} be convergent, it
is necessary that $p$ in Eq. \Ref{dp2} be even and $a_p$ be
positive.

For the case of large $N$, the summation can be rewritten in the
form of a path integral over continuous parameters. These
parameters are introduced as
\begin{align}\label{dp4}
0\leq x:=&\;\frac{i}{N}\leq 1,\nonumber\\
n(x):=&\;\frac{n_i}{N},\nonumber\\
\phi(x):=&\;-n(x)-1+x.
\end{align}
The partition function \Ref{dp1} then becomes
\begin{equation}\label{dp5}
Z=\int\prod_{0\leq x\leq 1}\d\phi(x)\;\exp[S(\phi)],
\end{equation}
where
\begin{equation}\label{dp6}
S(\phi):=N^2\,\left\{-A\int_0^1\d x\;G[\phi(x)]+\int_0^1\d
x\int_0^1 \d y\;\log|\phi(x)-\phi(y)|\right\},
\end{equation}
apart from an unimportant constant, and
\begin{equation}\label{dp7}
G(z):=\sum_{k=1}^{p}(-1)^k\,a_k\,z^k.
\end{equation}
Introducing the density
\begin{equation}\label{dp8}
\rho[\phi(x)]:=\frac{\d x}{\d\phi(x)},
\end{equation}
one would have
\begin{equation}\label{dp9}
\int_{b}^{a}\d z\;\rho(z)=1,
\end{equation}
(the so called normalization condition for the density), where $b$
and $a$ are the lower and upper limits of $\phi(x)$, respectively.
The condition that $n_i$'s be nonincreasing with respect to $i$
imposes the following condition on the density.
\begin{equation}\label{dp10}
0\leq\rho(z)\leq 1.
\end{equation}

The path integral in the right-hand side of \Ref{dp5} is dominated
by the classical representation, which maximizes the function $S$,
or minimizes the free energy $F$ defined through
\begin{equation}\label{dp11}
F:=-\frac{1}{N^2}\,\ln Z.
\end{equation}
Denoting the density corresponding to this representation by
$\rho_{\rc}$, it is seen that the derivative of the free energy
with respect to the area is
\begin{align}\label{dp12}
F'(A)=&\int_0^1\d x\;G[\phi_\rc(x)],\nonumber\\
=&\int_{b}^{a}\d z\;\rho_\rc(z)\,G(z),\nonumber\\
=&\oint_{C_\infty}\frac{\d z}{2\,\pi\,\ri}\,H_\rc(z)\,G(z),
\end{align}
where the function $H$ of a complex variable is defined through
\begin{equation}\label{dp13}
H(z):=\int_b^a\d\zeta\;\frac{\rho(\zeta)}{z-\zeta},
\end{equation}
and $C_\infty$ is a counterclockwise contour outside of which $H$
is analytic. It is seen from \Ref{dp9} that $H(z)$ behaves like
$z^{-1}$ for large $z$.

The classical density $\rho_\rc$ is the weak density $\rho_\rw$
satisfying
\begin{equation}\label{dp14}
g(z)=\mathrm{P}\int_{b_\rw}^{a_\rw}\d\zeta\;
\frac{\rho_{\mathrm{w}}(\zeta)}{z-\zeta},\quad b_\rw\leq z\leq
a_\rw,
\end{equation}
where P means principal value and
\begin{equation}\label{dp15}
g(z):=\frac{A}{2}\,G'(z),
\end{equation}
provided the conditions \Ref{dp10} are not violated by $\rho_\rw$.
This is the weak phase. One obtains
\begin{align}\label{dp16}
H_\rw(z)=&\;g(z)-\sqrt{(z-a_\rw)\,(z-b_\rw)}\nonumber\\
&\times \sum_{m,n,q=0}^{\infty}\frac{(2n-1)!!\,(2q-1)!!}{2^{n+q}\,
n!\,q!\,(n+q+m+1)!}\,a_\rw^n\,b_\rw^q\,z^m\,g^{(n+m+q+1)}(0),
\end{align}
where $g^{(n)}$ is the $n$'th derivative of $g$, and
\begin{align}\label{dp17}
\rho_\rw(z)=&\;\frac{\sqrt{(a_\rw-z)\,(z-b_\rw)}}{\pi}\nonumber\\
&\times\sum_{m,n,q=0}^{\infty}\frac{(2n-1)!!\,(2q-1)!!}{2^{n+q}\,
n!\,q!\,(n+q+m+1)!}\,a_\rw^n\,b_\rw^q\,z^m\,g^{(n+m+q+1)}(0).
\end{align}
The condition that $H(z)$ behaves like $z^{-1}$ for large $z$, is
equivalent to the following equations.
\begin{align} \label{dp18}
\sum_{n,q=0}^{\infty}\frac{(2n-1)!!\,(2q-1)!!}{2^{n+q}\,n!\,q!\,
(n+q)!}\,a_\rw^n\,b_\rw^q\,g^{(n+q)}(0)=&\;0,\\ \label{dp19}
\sum_{n,q=0}^{\infty}\frac{(2n-1)!!\,(2q-1)!!}{2^{n+q}\,n!\,q!\,
(n+q-1)!}\,a_\rw^n\,b_\rw^q\,g^{(n+q-1)}(0)=&\;1.
\end{align}
These equations are used to obtain $a_\rw$ and $b_\rw$.

Now consider a function $G$ like
\begin{equation}\label{dp20}
G(z)=z^4+\lambda\,z^3.
\end{equation}
The conditions \Ref{dp18} and \Ref{dp19} become
\begin{align}\label{dp21}
\tau_\rw^2\,\left(3\,\sigma_\rw+\frac{3}{4}\,\lambda\right)+
\sigma_\rw^2\,\left(2\,\sigma_\rw+\frac{3}{2}\,\lambda\right)=&\;0,
\\ \label{dp22} \frac{3}{4}\,\tau_\rw^4+\tau_\rw^2\,
\left(3\,\sigma_\rw^2+\frac{3}{2}\,\sigma_\rw\,\lambda\right)=&\;\frac{1}{A},
\end{align}
where
\begin{align}\label{dp23}
\sigma:=&\;\frac{a+b}{2},\nonumber\\
\tau:=&\;\frac{a-b}{2}.
\end{align}
Using \Ref{dp16}, one obtains
\begin{align}\label{dp24}
H_\rw(z)=\frac{A}{2}\,&\{4\,z^3+(3\,z^2)\,\lambda-
\sqrt{(z-\sigma_\rw)^2-\tau_\rw^2}\nonumber\\
&\times[4\,\sigma_\rw^2 +2\,\tau_\rw^2+4\,\sigma_\rw\,z+4\,z^2+
3\,(\sigma_\rw+z)\,\lambda]\}.
\end{align}
The solution to \Ref{dp21} and \Ref{dp22} for small values of
$\lambda$ is
\begin{align}\label{dp25}
\sigma_\rw=&\;-\frac{1}{4}\,\lambda-\frac{(3\,A)^{1/2}}{96}\,
\lambda^3+O(\lambda^5),\nonumber\\
\tau_\rw^2=&\;\frac{2}{(3\,A)^{1/2}}+\frac{1}{8}\,\lambda^2+O(\lambda^4).
\end{align}
Using these, \Ref{dp12}, and \Ref{dp24}, one obtains a
perturbative expression for the derivative of the free energy with
respect to the area:
\begin{equation}\label{dp26}
F_\rw'(A)=\frac{1}{4\,A}-\frac{1}{8\,(3\,A)^{1/2}}\,\lambda^2+O(\lambda^4).
\end{equation}

To investigate the conditions \Ref{dp10} for $\rho_\rw$, one
obtains $\rho_\rw$ and its minima and maxima. From \Ref{dp17}, one
has
\begin{equation}\label{dp27}
\rho_\rw(z)=\frac{A}{2\,\pi}\,\sqrt{\tau_\rw^2-(z-\sigma_\rw)^2}\,
[4\,\sigma_\rw^2 +2\,\tau_\rw^2+4\,\sigma_\rw\,z+4\,z^2+
3\,(\sigma_\rw+z)\,\lambda].
\end{equation}
Using this, one obtains three points where the derivative of
$\rho_\rw$ vanishes:
\begin{align}\label{dp28}
z_0=&\;-\frac{1}{4}\,\lambda+\frac{(3\,A)^{1/2}}{96}\,\lambda^3+O(\lambda^5),
\nonumber\\
z_{\pm}=&\;\pm\frac{1}{(3\,A)^{1/4}}-\frac{1}{4}\,\lambda
\pm\frac{(3\,A)^{1/4}}{16}\,\lambda^2-\frac{(3\,A)^{1/2}}{96}\,\lambda^3
+O(\lambda^4),
\end{align}
and the corresponding values for the density $\rho_\rw$:
\begin{align}\label{dp29}
\rho_\rw(z_0)=&\;\frac{\sqrt{8}\,(3\,A)^{1/4}}{3\,\pi}-
\frac{(3\,A)^{3/4}}{\sqrt{128}\,\pi}\,\lambda^2+O(\lambda^4),
\nonumber\\
\rho_\rw(z_{\pm})=&\;\frac{4\,(3\,A)^{1/4}}{3\,\pi}
\mp\frac{A}{16\,\pi}\,\lambda^3+O(\lambda^5).
\end{align}
The absolute maximum of $\rho_\rw$ exceeds one at $A=A_\c$. This
gives the critical area $A_\c$:
\begin{equation}\label{dp30}
A_\c=A_\c(0)\,\left[1-\frac{A_\c(0)}{4\,\pi}\,|\lambda|^3
+O(|\lambda|^5)\right],
\end{equation}
where
\begin{equation}\label{dp31}
A_\c(0)=\frac{27\,\pi^4}{256}.
\end{equation}
It is seen that in the limit $\lambda=0$ one recovers the result
obtained in \cite{20}.

One can also obtain the critical exponent relating $\lambda$ and
$A_\c$. For a specific value of $A_\c$, one obtains two values for
$\lambda$. Denoting the difference of these values by
$\Delta\lambda$, it is seen that
\begin{equation}\label{dp32}
\Delta\lambda\sim[A_\c(0)-A_\c]^{1/3}.
\end{equation}

\section{The strong phase}
For $A>A_\c$, there is a region where the density $\rho_\rw$
exceeds one. Then $\rho_\rc$ cannot be $\rho_\rw$ and one uses
another ansatz for $\rho_\rc$, which is $\rho_\rs$:
\begin{equation}\label{dp33}
\rho_\rs(z)=\begin{cases} 1,& z\in[d_\rs,c_\rs]\\
\tilde\rho_\rs(z),& z\in[b_\rs,d_\rs]\cup[c_\rs,a_\rs]
\end{cases}.
\end{equation}
This is the strong phase. An argument similar to that used in
\cite{20} shows that
\begin{align}\label{dp34}
H_\rs(z)=\frac{A}{2}\,&\{4\,z^3+(3\,z^2)\,\lambda
\nonumber\\
&-\sqrt{[(z-\sigma_\rs)^2-\tau_\rs^2]\,[(z-s_\rs)^2-t_\rs^2]}\,
[4\,(\sigma_\rs+s_\rs+z)+3\,\lambda]\}\nonumber\\
&-\sqrt{[(z-\sigma_\rs)^2-\tau_\rs^2]\,[(z-s_\rs)^2-t_\rs^2]}\,
\int_{d_\rs}^{c_\rs}\frac{\d\zeta}{(z-\zeta)\,R(\zeta)},
\end{align}
where
\begin{align}\label{dp35}
s:=&\;\frac{c+d}{2},\nonumber\\
t:=&\;\frac{c-d}{2},\nonumber\\
R(z):=&\;\sqrt{[\tau_\rs^2-(z-\sigma_\rs)^2]\,[t_\rs^2-(z-s_\rs)^2]}.
\end{align}
$H_\rs(z)$ should behave like $z^{-1}$ for $z\to\infty$. Expanding
$H_\rs(z)$ for large $z$, It is seen that
\begin{equation}\label{dp36}
H_\rs(z)=\alpha_1\,z+\alpha_0+\alpha_{-1}\,z^{-1}+O(z^{-2}).
\end{equation}
Hence one arrives at three equations
\begin{align}\label{dp37}
\alpha_1=&\;0,\nonumber\\
\alpha_0=&\;0,\nonumber\\
\alpha_{-1}=&\;1.
\end{align}
One also has
\begin{equation}\label{dp38}
\int_{d_\rs}^{c_\rs}\d z\;[g(z)-H_\rs(z)]=0,
\end{equation}
which is
\begin{equation}\label{dp39}
\frac{A}{2}\,\int_{d_\rs}^{c_\rs}\d
z\;[4\,(\sigma_\rs+s_\rs+z)+3\,\lambda]\,R(z)+
\mathrm{P}\int_{d_\rs}^{c_\rs}\d z\;\int_{d_\rs}^{c_\rs}\d\zeta\;
\frac{R(z)}{R(\zeta)}\,\frac{1}{z-\zeta}=0.
\end{equation}
Using this equation and the equations \Ref{dp37}, one can obtain
the four unknowns $\sigma_\rs$, $\tau_\rs$, $s_\rs$, and $t_\rs$.
Expanding these in terms of $(A-A_\c)$, and using \Ref{dp12}, one
can obtain an expansion for the derivative of the free energy with
respect to the area.

However, to obtain the behavior of the free energy in the strong
phase at the limit $\lambda\to 0$, there exists a simpler way
based on \cite{20} and \cite{l}. It is known from \cite{l} that
the transition is of third order, meaning that the difference
between $H_\rs(z)$ and $H_\rw(z)$ for large $z$ behaves like
$(A-A_\c)^2$:
\begin{equation}\label{dp40}
\lim_{A\to
A^+_\c(\lambda)}\{[A-A_\c(\lambda)]^{-2}\,[H_\rs(z)-H_\rw(z)]\}=B\left[z,\lim_{A\to
A_\c(\lambda)}\rho_\rw\right].
\end{equation}
Increasing $A$ further, $\rho_\rs$ itself would exceeds one in
some region. That is, a second phase transition occurs. Denoting
the classical density after this transition by $\rho_{\rs\,2}$,
and the area corresponding to this transition by $A_{\c\,2}$, one
has
\begin{equation}\label{dp41}
\lim_{A\to A^+_{\c\,2}(\lambda)}\{[A-A_{\c\,2}(\lambda)]^{-2}\,
[H_{\rs\,2}(z)-H_\rs(z)]\}=B\left[z,\lim_{A\to
A_{\c\,2}(\lambda)}\rho_\rs\right].
\end{equation}
Now consider the limit $\lambda\to 0$. At this limit the two
transition occur at the same area $A_\c(0)$. One also has
\begin{equation}\label{dp42}
\lim_{\lambda\to 0}\left[\lim_{A\to
A_{\c\,2}(\lambda)}\rho_\rs\right]=\lim_{\lambda\to
0}\left[\lim_{A\to A_\c(\lambda)}\rho_\rw\right].
\end{equation}
So, adding \Ref{dp40} and \Ref{dp41} at the limit $\lambda\to 0$,
one arrives at
\begin{equation}\label{dp43}
\lim_{A\to A^+_\c(0)}\{[A-A_\c(0)]^2\,\lim_{\lambda\to
0}[H_{\rs\,2}(z)-H_\rw(z)]\}= 2\,B\left[z,\lim_{\lambda\to
0}\lim_{A\to A_\c(0)}\rho_\rw\right].
\end{equation}
Defining
\begin{equation}\label{dp44}
I:=\;\lim_{A\to A^+_\c(0)}\{[A-A_\c(0)]^2\,\lim_{\lambda\to
0}[F'_{\rs\,2}(A)-F'_\rw(A)]\},
\end{equation}
it is then seen from \Ref{dp40} and \Ref{dp43} that
\begin{equation}\label{dp45}
\lim_{A\to A^+_\c(0)}\{[A-A_\c(0)]^2\,\lim_{\lambda\to
0}[F'_\rs(A)-F'_\rw(A)]\}=\frac{1}{2}\,I,
\end{equation}
where \Ref{dp12} has been used the relate $F'$ to $H$.

Now consider the case $\lambda=0$, which has been investigated in
\cite{20}. When $\lambda=0$, the density $\rho_\rw$ has two equal
maxima and at the transition both maxima exceed one. It is then
seen that $[\lim_{\lambda\to 0}F'_{\rs\,2}(A)]$ is in fact the
derivative of the free energy in the strong phase for the case
$\lambda=0$ studied in \cite{20}. So the difference between
$F'_\rs(A)$ and $F'_\rw(A)$ in the limit $\lambda\to 0$, is half
what obtained in \cite{20}:
\begin{equation}\label{dp46}
\lim_{\lambda\to 0}[F'_\rs(A)-F'_\rw(A)]=\frac{1}{54\,A_\c(0)}\,
\left(\frac{A-A_\c}{A_\c}\right)^2+
O\left[\left(\frac{A-A_\c}{A_\c}\right)^3\right].
\end{equation}

\section{The second transition}
As it was pointed out in the previous section, $\rho_\rw$ has two
maxima which are equal when $\lambda=0$. If $\lambda$ is not zero,
then one of these becomes larger and the absolute maximum of
$\rho_\rw$. The transition point is where this maximum is going to
exceed one. After this transition, the classical density is
$\rho_\rs$, for which there is a region where the density is equal
to one. However, in the region around the point the other maximum
of $\rho_\rw$ occurs, the density $\rho_\rs$ continues to increase
and there may be some point where this second maximum exceeds one
as well. Here the second transition occurs. For small values of
$\lambda$ it is easy to find this point. Once again the result of
\cite{l} is used. There it is proved that the difference between
the $\rho_\rs(z)$ and $\rho_\rw(z)$ vanishes faster than
$(A-A_\c)$, provided the distance of $z$ from the region where
$\rho_\rs$ is one is large compared to the width of the region.
This criterion is satisfied for $z$ around the point where the
second maximum of $\rho_\rw$ occurs. So,
\begin{equation}\label{47}
\rho_\rs(z)=\rho_\rw(z)+o[(A-A_\c)],\quad z\approx z_\m,
\end{equation}
where $z_\m$ is the point where the second maximum of $\rho_\rw$
occurs. This shows that up to leading order, the second transition
occurs where the second maximum of $\rho_\rw$ approaches one. So,
\begin{equation}\label{dp48}
A_{\c\,2}=A_\c(0)\,\left[1+\frac{A_\c(0)}{4\,\pi}\,|\lambda|^3\right]+\cdots
\end{equation}
The phase picture of the system for small $\lambda$ is now
complete. There are three phases: weak, strong, and stronger.
\begin{itemize}
\item[$\bullet$] \textbf{The weak phase.} Here one has
\begin{equation}\label{dp49}
A<A_\c(0)\,\left[1-\frac{A_\c(0)}{4\,\pi}\,|\lambda|^3\right],
\end{equation}
the classical density is $\rho_\rw$, and the classical density is
everywhere less than one.
\item[$\bullet$] \textbf{The strong phase.} Here one has
\begin{equation}\label{dp50}
A_\c(0)\,\left[1-\frac{A_\c(0)}{4\,\pi}\,|\lambda|^3\right]<
A<A_\c(0)\,\left[1+\frac{A_\c(0)}{4\,\pi}\,|\lambda|^3\right],
\end{equation}
the classical density is $\rho_\rs$, and there is one interval
where the classical density is equal to one. For fixed $\lambda$,
the area interval corresponding to this phase is seen to be
proportional to $\lambda^3$. This phase consists of two parts. For
positive (negative) $\lambda$ the region where $\rho_\rc$ is one
is around $z_-$ ($z_+$).
\item[$\bullet$] \textbf{The stronger phase.} Here one has
\begin{equation}\label{dp51}
A_\c(0)\,\left[1+\frac{A_\c(0)}{4\,\pi}\,|\lambda|^3\right]<A,
\end{equation}
the classical density is $\rho_{\rs\,2}$, and there are two
intervals where the classical density is equal to one.
\end{itemize}
It is seen that if $\lambda$ vanishes, the strong phase disappears
and only two phases remain, which agrees with the result of
\cite{20}.

\section{Concluding remarks}
A gYM$_2$ with quartic and cubic couplings was studied. The effect
of the cubic coupling on the classical density and the free energy
was investigated. The quantitative changes in the weak phase were
determined for small cubic couplings. The strong phases of the
system were also investigated. It was seen that for small but
nonvanishing cubic couplings there are two transitions, compared
to a single transition if there is no cubic coupling. Both of
these transitions are third order (in area). It was also seen that
for small but nonvanishing cubic couplings, the jump in the third
derivative of the free energy in each of these transitions is half
the jump corresponding to the single transition which occurs if
there is no cubic coupling.\\
\\
\textbf{Acknowledgement}\\
This work was partially supported by the research council of the
Alzahra University.

\end{document}